\documentclass[%
 reprint,
superscriptaddress,
onecolumn,
 amsmath,amssymb,
 aps,
]{revtex4-2}
\usepackage{adjustbox}
\usepackage{physics}
\usepackage{graphicx}
\usepackage{dcolumn}
\usepackage{bm}

\usepackage{chemformula} 
\usepackage{booktabs}
\usepackage{graphicx}
\newcommand{\cpo}{\alpha}
\newcommand{\Vic}{V_i^{cap}}

\begin{document}

\preprint{APS/123-QED}

\title{Breaking one into three: surface-tension-driven droplet breakup in T-junctions}

\author{Jiande Zhou}
\affiliation{%
 Laboratory of Microsystems (LMIS4), EPFL, CH1015 Lausanne, Switzerland
}%
\author{Yves-Marie Ducimetière}
\affiliation{%
 Laboratory of Fluid Mechanics and Instabilities, EPFL, CH1015 Lausanne, Switzerland
}%
\author{Daniel Migliozzi}
\affiliation{%
 Laboratory of Microsystems (LMIS4), EPFL, CH1015 Lausanne, Switzerland
}
 \author{Ludovic Keiser}
\affiliation{%
 Laboratory of Fluid Mechanics and Instabilities, EPFL, CH1015 Lausanne, Switzerland
}
 \affiliation{Univ. Grenoble Alpes, CNRS, LIPhy, 38000 Grenoble, France}
 
\author{Arnaud Bertsch}
\affiliation{%
 Laboratory of Microsystems (LMIS4), EPFL, CH1015 Lausanne, Switzerland
}%
\author{François Gallaire}
\affiliation{%
 Laboratory of Fluid Mechanics and Instabilities, EPFL, CH1015 Lausanne, Switzerland
}%
\author{Philippe Renaud}
\affiliation{%
 Laboratory of Microsystems (LMIS4), EPFL, CH1015 Lausanne, Switzerland
}%





\date{\today}

\begin{abstract}

Droplet breakup is an important phenomenon in the field of microfluidics to generate daughter droplets. In this work, a novel breakup regime in the widely studied T-junction geometry is reported, where the pinch-off occurs laterally in the two outlet channels, leading to the formation of three daughter droplets, rather than at the center of the junction for conventional T-junctions which leads to two daughter droplets. It is demonstrated that this new mechanism is driven by surface tension, and a design rule for the T-junction geometry is proposed. A model for low values of the capillary number $Ca$ is developed to predict the formation and growth of an underlying carrier fluid pocket that accounts for this lateral breakup mechanism. At higher values of $Ca$, the conventional regime of central breakup becomes dominant again. The competition between the new and the conventional regime is explored. Altogether, this novel droplet formation method at T-junction provides the functionality of alternating droplet size and composition, which can be important for the design of new microfluidic tools.

\end{abstract}

\maketitle

\section{Introduction}

Droplet formation is a ubiquitous process in both nature and industry. In the context of microfluidics, the controllable generation of micro-droplets has enabled a wide range of applications, opening a new era for biological and chemical analysis and synthesis \cite{shang2017emerging,zhu2017passive,joanicot2005droplet}. The formation of droplets is the first step to achieve in the pipeline in order to achieve versatile functionalities such as microreactors \cite{sun2020droplet,swank2021high,wang2021high}, mini-incubators \cite{kulesa2018combinatorial,schuster2020automated,wang2021high}, material templates \cite{durmus2013functional,zhang2020continuous,kumar2021droplet}, digital counters \cite{hindson2013absolute,zhang2019accurate,zaremba2021integration}, or single cell platforms \cite{segaliny2018functional,pellegrino2018high,spindler2020massively,gerard2020high}. To date, droplet formation mechanisms in rectangular microchannels have been widely studied and can be classified in two main categories \cite{thorsen2001dynamic,anna2003formation,cramer2004drop,xu2008correlations,baroud2010dynamics,korczyk2019accounting}:  the mechanisms driven by hydrodynamic forces and those driven by surface tension. In the former category, the carrier flow is brought to the dispersed phase to generate viscous and/or inertial forces destabilizing the interface, and surface tension acts as the stabilizing force. In the latter, the interface breakup is purely driven by an imbalance in capillary pressure induced by an abrupt change of confinement \cite{dangla2013physical}. The first category of droplet production processes is flexible in operation and advantageous in producing a high droplet throughput \cite{xu2008correlations,de2008transition,wang2021high}. Although limited by the flow rate, the second category is advantageous for monodisperse droplet production and  parallelization \cite{dangla2013droplet,li2015step,Eggersdorfer2018}. 

Droplet formation can result from the emulsification of a continuous phase, or from the breakup of an existing droplet. The latter process enables to adjust the initial droplet size, to increase droplet production rate or to provide new functionalities, such as up-concentration \cite{niu2011microdroplet,lan2016droplet}. One of the most studied geometry for droplet breakup is the T-junction, where a straight channel splits perpendicularly into two lateral channels. 
Following the seminal work by Link. \textit{et al.} \cite{Link2004}, various studies have investigated the dynamics of the droplet breakup process, for both short \cite{de2006modeling,Leshansky2009,afkhami2011numerical,chen20153d,chen2017hydrodynamics,sun2018dynamics,sun2019breakup} and elongated droplets \cite{Jullien2009,Hoang2013,leshansky2012,wang2015dynamics,Haringa2019,chang2019dynamics,mora2019numerical}. Other studies also investigated how to modify the topology of the T-junction to perform asymmetric droplet breakup \cite{Samie2013,Bedram2015,Zheng2016}. In all those configurations, droplets don't breakup at small capillary numbers $Ca = \mu V /\gamma $, and are split into two daughter droplets above a critical capillary number $Ca_c = \mu V_c / \gamma$, with $\mu$ the viscosity of the carried fluid, $\gamma$ the interfacial tension and $V$ the speed of the droplet. The breakup process is here driven by the hydrodynamics stress exerted by the carrier flow which enables to deform and break the interface. 

In this study, we report a novel droplet break-up regime in T-junctions that is surface-tension-driven. In this regime, the droplet interface ruptures symmetrically in the two lateral channels away from the junction, which gives birth to three daughter droplets instead of two. We show that this regime only occurs in T-junctions that have a different aspect and width ratio compared to the ones presented so far in the scientific literature until now. The height $h$ of the channels must be larger than the width of the inlet channel $w_i$, which itself must be larger than the width of the outlet channel $w_o$: $h>w_i>w_o$. We describe the underlying mechanism of the new droplet breakup mechanism and provide a geometry design rule predicting the occurrence of the new regime in a T-junction. We also propose a semi-quantitative model accounting for the gutter flows to describe the dynamical process of the new breakup regime. Finally, we show that the conventional central breakup also occurs in the new T-junctions under certain flow conditions. Both central and lateral breakup regimes can develop independently, but the droplet breakup regime actually occurring is the faster one.

\begin{figure*}
    \centering
	\includegraphics[scale=0.9]{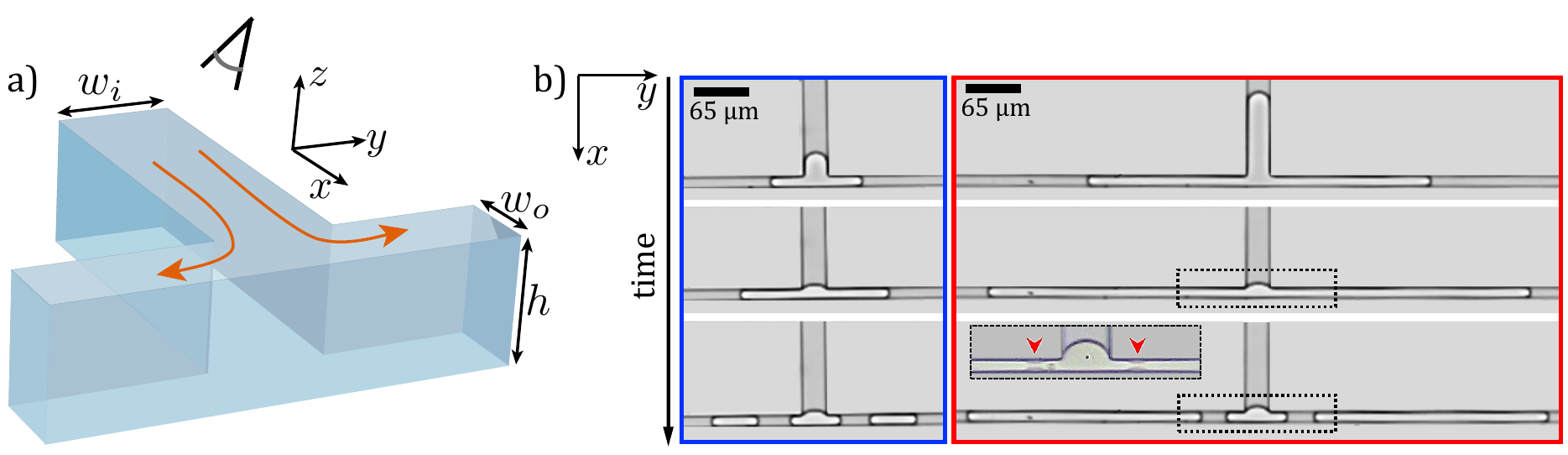}
  \caption{\textbf{(a)} Geometry of the T-junctions used throughout our study: both aspect ratio $h/w_o$ and width ratio $w_i/w_o$ are larger than unity. The eye shows the observation perspective during experiment.\textbf{(b)} Time sequence of a lateral breakup process for a short (blue) and a long (red) droplet. Inset shows the interface at the moment of rupture (red arrow), captured at a frame rate of 50,000 fps.}
  \label{fgr:intro}
\end{figure*}

\section{Experimental and numerical methods}
\subsection{Device fabrication}
To create the microchannels, a silicon mold fabricated by Dry Reactive Ion Etching (DRIE) was used. First, a 1.5 $\mu$m photoresist layer was deposited on double side polished silicon wafers, and was patterned with standard photolithography including steps of exposure and development to obtain the 2D channel shape. The exposed wafer area was then etched using the Bosch process (DRIE, Alcatel AMS 200). The obtained channel depth is proportional to the etching duration, and the value is measured with a surface profilometer (Tencor Alpha-Step 500). After the Si mold was silanized within a trichlor-(1H,1H,2H,2H-perfluoroctyl) (PFOTs)-filled desiccator for five hours, we pour PDMS pre-polymer (1:10 ratio mixture) onto the Si mold and cure in 80\textdegree oven for three hours. We peel the PDMS replicas from the mold and after punching inlet and outlet holes seal the channels by bonding to a PDMS-coated glass slide (oxygen plasma bonding, 500 mTorr, 45 sec, 29 W). Coating of the glass slide (standard 25mm * 75mm * 1mm) is done by spin coating a thin layer of PDMS pre-polymer at 1700 rpm for 35s, then cure in the oven (as above).The hydrophobicity of the surfaces was naturally regained by placing PDMS in the oven for 3 days.

\subsection{Experiments}
The experiments were performed under an inverted microscope (Nikon Eclipse TE 300) and imaged with a high-speed camera (Phantom Miro M310). Syringe pump (CETONI Nemesys) with gastight glass syringes (Hamilton) is used to control the flowrates injected into the system. Depending on the flow rates, frame rates up to 50,000 frames per second were used for recording the droplet breakup process. A customized ImageJ script is used to automatically recognize the droplets and obtain the intensity profile. A MATLAB (Mathworks) script is used for calculating the droplet speed and length. The results were confronted to the observations to ensure accuracy.

\subsection{Numerical simulations}
The closed system for the three unknowns $\bar{z}(\eta,t)$, $\bar{k}(\eta,t)$ and $L_p(t)$ from equations (\ref{eq:sysnv}) and (\ref{eq:forLp})  was solved numerically using the COMSOL Multiphysics software, based on the finite element method. More precisely, the system (\ref{eq:sysnv}) is implemented directly in the "General form PDE" component of the software, and it weak form has been spatially discretized over the interval $\xi \in [0,1]$ using first-order polynomials (corresponding to a linear interpolation of the solution). The convergence of the numerical results with respect to the spatial discretization has been verified. Equation (\ref{eq:forLp}), enforcing the volume conservation thus depending only on time, is implemented in the "Global Equations" component. The system is marched in time using the backward differentiation formula ("BDF"), and the results are sought for $4000$ discrete times uniformly distributed between $0$ and $10/Ca$. A stopping condition has been added, such that the simulation stops running if $\min(\bar{z})<1/2$. Concerning the nonlinear, fully coupled solver, the default choices of COMSOL parameters have been found sufficient for convergence, excepted the "Jacobian update" that is set to "updated on every iteration" and the "Maximum number of itertions" that is set to $500$.

\section{Results}

\subsection{Description of a novel breakup mechanism}

In previous studies, droplet breakup was conducted in T-junctions, where the inlet and outlet channels have the same width and where the channel height is equal to or smaller than the width \cite{Link2004,Leshansky2009,Haringa2019}. 
In this study, we use a non-conventional T-junction (\textbf{Figure \ref{fgr:intro}a}) with the inlet and outlet width ($w_i$ and $w_o$ respectively) and the height of the channel ($h$) fulfilling $h> w_i > w_o$. Consequently, both the aspect ratio ($h/w_o$) and the width ratio ($w_i/w_o$) are larger than unity. Upstream of the T-junction, water-in-oil droplets are generated using a flow-focusing device with two inlets, one introducing deionized water (dispersed phase) and the other fluorinated oil (continuous phase). A third inlet introduces additional oil downstream of the flow-focusing unit in order to separate the droplets and further control their speed. When a droplet passes through the T-junction and fully enters the lateral channels, its rear interface remains pinned at the junction with a constant and convex curvature, whereas the front interfaces advance further downstream. This is in contrast with the  central breakup mechanism which features a progressive concave curving of the rear interface during breakup  \cite{Link2004}. Eventually, it is the interfaces inside the lateral channels that collapse and create two new interfaces at a symmetric distance from the junction. This type of breakup creates three daughter droplets rather than two (which is observed during the classical breakup at T-junction). This breakup is referred to as \textit{lateral breakup}, in comparison to the classical \textit{central breakup} described in the literature. Two examples of droplets undergoing a lateral breakup are presented in \textbf{Figure \ref{fgr:intro}b}. The collapse is very rapid but can be captured by a high speed camera (inset).

 The interface appears to break very suddenly during the time the rear cap remains pinned at the junction. It suggests that the necking process, which is usually more gradual, is likely acting off-plane before the final pinch-off happens.

\subsection{Geometric conditions required for the lateral breakup}
We first detail the geometry of the T-junction which enables the lateral breakup phenomenon. Using a quasi-static assumption, we consider that the Young-Laplace equation controls the pressure drop across the interface of the droplet: $ p_d-p= \gamma \kappa$, where $p_d$ is the pressure in the drop, $p$ is the pressure in the surrounding fluid at the interface (\textbf{Figure \ref{fgr:static}a}), $\gamma$ is the interfacial tension, and $\kappa$ is the local mean curvature of the interface. When the droplet passes through a T-junction whose outlet channels have smaller dimension than the inlet channel ($w_o<w_i$), the curvature at the front of the droplet $\kappa_o$ increases compared to the curvature at the rear $\kappa_i$, thereby creating a pressure gradient along the droplet interface. 
We assume that the dominant pressure variations occur in the gutters present in the corners of the cross-section, and consider that the pressure in the droplet $p_d$ can be roughly considered as constant in space and time \cite{Haringa2019,van2013block}. It implies that in the fluid surrounding the droplet, the pressure gradually decreases  from the rear cap in the inlet channel ($p_i=p_d-\gamma \kappa_i$) to the front cap in the outlet channels ($p_o=p_d-\gamma \kappa_o$), by continuity.
This pressure gradient is accompanied by an adaptation of the radius of the gutter $R_g$, such that $p=p_d - \gamma/R_g$. Along the droplet, $R_g$ thus varies from $1/\kappa_i$ in the rear of the droplet to $1/\kappa_o$ in the front.
In the quasi-static condition, the value of $\kappa_i$ and $\kappa_o$ is constant and only dependent on the channel geometry\cite{wong1992three}:

\begin{figure*}
	\includegraphics[scale=0.8]{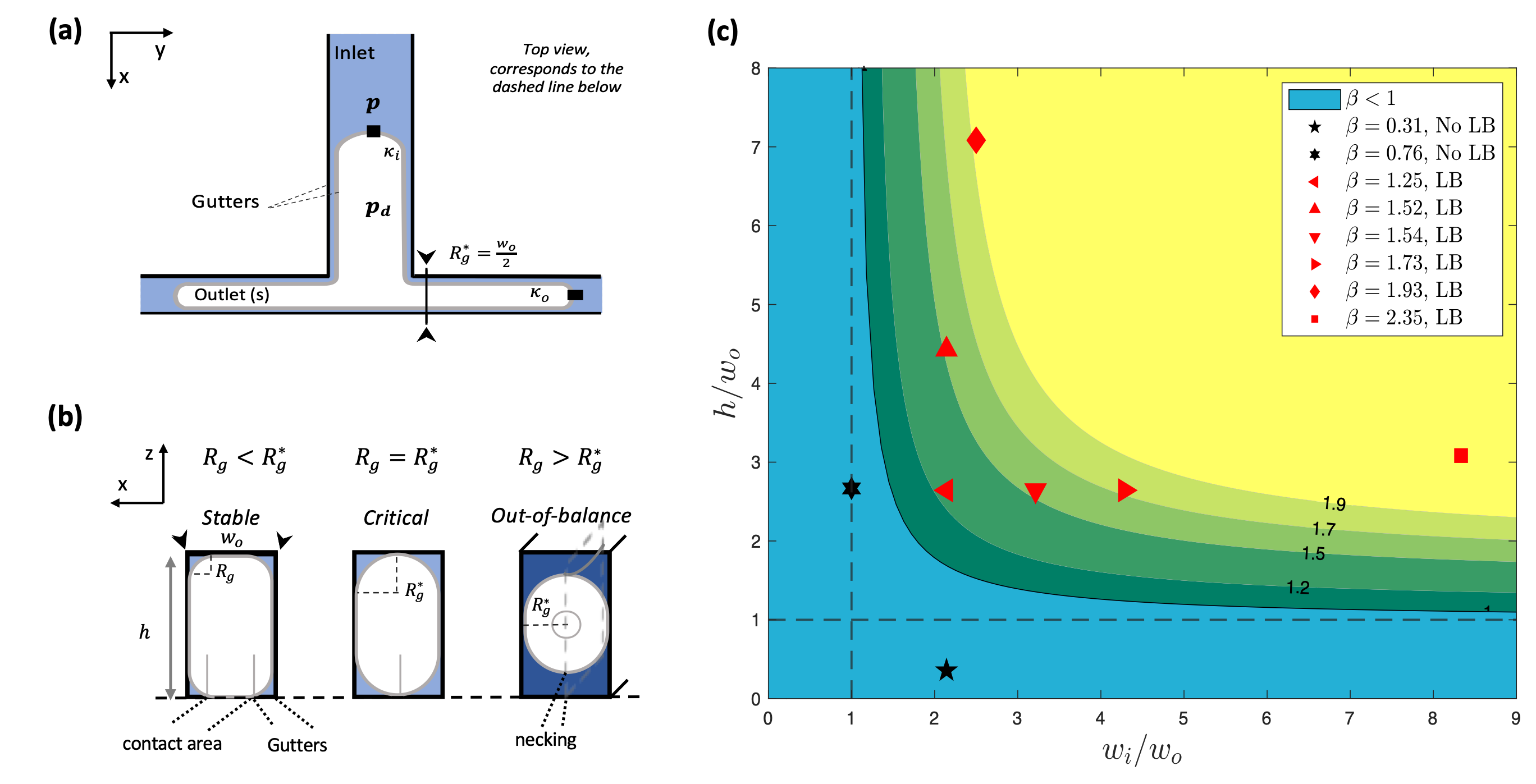}
  \caption{Capillary instability responsible for lateral breakup (a)Presentation of key droplet parameters with a top view of a T-junction through which a droplet splits. (b) Cross sectional view in the outlet channel (indicated by black arrowheads in (a)) showing three different gutter radius. These three conditions can hold for the same droplet passing the junction at different times. When $R_g >R_g^*$, the necking starts. (c) Colormap representing $\beta$ in a diagram representing the aspect ratio $h/w_o$ as a function of the width ratio $w_i/w_o$. Blue region corresponds to $\beta<1$, i.e. geometries that does \textit{not} allow the occurrence of lateral break-up; Green regions correspond to $\beta>1$, i.e. geometries prone to exhibit lateral breakup. Each marker corresponds to one of the geometries experimentally tested (see Table.\ref{tab:geo}). A black symbol correspond to an absence of lateral breakup observed, and a red marker corresponds to a presence of a lateral breakup.}
  \label{fgr:static}
\end{figure*} 

\begin{equation}
\kappa_{i,o} = \frac{1+ \frac{w_{i,o}}{h}+\sqrt{\left[(1-\frac{w_{i,o}}{h})^2+\pi\frac{w_{i,o}}{h}\right]}}{w_{i,o}}.
\label{eq:kappa}
\end{equation}
where \textit{i} and \textit{o} respectively account for inlet and outlet. However due to the confinement and the non-wetting condition, $R_g$ cannot exceed a threshold value given by half of the smallest dimension of the cross-section \cite{dangla2013droplet}. In our case $w_o<h$ which gives a critical value $R_g^*$ in the outlet channel : $R_g^*=w_o/2$. Consequently, if the relatively large pressure imposed in the gutter by the proximity of the rear cap imposes a radius of curvature $R_g$ larger than $R_g^*$ in the outlet channels, an instability is triggered (\textbf{Figure \ref{fgr:static}b}). In order to fulfill the continuity of pressure along the channel, the interface has to curve concavely in the $y$ direction, which marks the initiation of a necking process.
A ``pocket'' thus gradually inflates at the upper and bottom part of the channel between the droplet and the ($x$,$y$)-walls, where the continuous phase accumulates. The droplet thereby thins down (necking process) until reaching a quasi-cylindrical shape, when surface tension induces a final and sudden breakup. Such a necking process is off-plane until the last moment of rupture, which is in consistence with the experimental observation. The lateral breakup most often occurs simultaneously in both outlet channels. 

\begin{table}[ht]
\centering
\caption{Tested geometries and the corresponding breakup outcome.}
\label{tab:geo}
\resizebox{0.85\textwidth}{!}{%
\begin{tabular}{@{}llllllllll@{}}
\toprule
Geometry & $w_i$ [$\mu m$] & $w_o$ [$\mu m$] & $h$ [$\mu m$] & $h/w_{o}$ & $w_{i}/w_{o}$ & $k_{i}$ & $\beta$ & Ca range & Lateral breakup \\ \midrule
A & 30  & 14 & 62 & 4.2 & 2.1 & 0.09 & 1.5 & 0.005-0.30  & Yes   \\
B & 30  & 14 & 37 & 2.6 & 2.1 & 0.11 & 1.2 & 0.006-0.077 & Yes   \\
C & 30  & 14 & 5  & 0.4 & 2.1 & 0.45 & 0.3 & 0.006-0.107 & No    \\
D & 60  & 14 & 37 & 2.6 & 4.3 & 0.08 & 1.7 & 0.004-0.183 & Yes   \\
E & 45  & 14 & 37 & 2.6 & 3.2 & 0.09 & 1.5 & 0.014-0.138 & Yes   \\
F & 30  & 30 & 80 & 2.8 & 1   & 0.09 & 0.8 & 0.006-0.160 & No    \\
G & 100 & 12 & 37 & 3.1 & 8.3 & 0.07 & 2.4 & 0.035-0.73  & Yes  \\
H & 30  & 12 & 85 & 7.1 & 2.5 & 0.09 & 1.9 & 0.02-0.04   & Yes \\ \bottomrule
\end{tabular}%
}
\end{table}

Note that the necking criterion $R_g>R_g^*=w_o/2$ is analogous to the one ruling step emulsification \cite{dangla2013physical} or snap-off \cite{Ransohoff88} processes.
The geometric criterion for a T-junction to allow a passing droplet meet the necking condition of $R_g>R_g^*$ can be expressed as $1/\kappa_i > R_g^*$. Defining the confinement parameter $\beta$ as $\beta = 2/(\kappa_iw_o)$, the lateral breakup will thus be prone to happen when $\beta > 1$, i.e. when:
\begin{equation}
\beta=2\frac{w_i/w_o}{1+ \frac{w_{i}}{h}+\sqrt{\left[(1-\frac{w_{i}}{h})^2+\pi\frac{w_{i}}{h}\right]}}>1.
\label{eq:beta}
\end{equation}
To test the criterion of Eq.\ref{eq:beta}, we select eight T-junction geometries with different combinations of $h/w_o$ and $w_i/w_o$ on which droplet breakup experiments are conducted with varying $Ca$ (see Table.\ref{tab:geo}). The outcomes of those experiments are represented in  \textbf{Figure \ref{fgr:static}c)}. No lateral breakup was observed for the two geometries with $\beta<1$, while geometries that have $\beta>1$ always showed lateral breakup for a given range of $Ca$ and $L$. In addition, with extreme $\beta$ value the lateral breakup phenomenon is significantly enhanced (more details in (Appendix)). Both observations confirm $\beta$ as a good proxy to evaluate the 'proneness' of the lateral breakup. We rearrange equation (\ref{eq:beta}) and map in fig.\ref{fgr:static}c.(ii) the $\beta$ value of a geometry as a function of its width ratio $w_i/w_o$ and aspect ratio $h/w_o$, where the green to yellow region represents the geometrical conditions of $\beta > 1$, which should allow the lateral breakup to occur. Note that a higher $\beta$ value is always associated with a larger aspect ratio and/or a larger width ratio, shown in the contour map as markers that are further and further away from the diagonal. It suggests that the capillary instability leading to the lateral breakup is driven by both of the two ratios. Indeed, the high aspect ratio ($h>w_o, h>w_i$) ensures that the confinement level on a droplet is dictated by the channel width (the smaller dimension). Then, the large width ratio ($w_i>w_o$) actually imposes the difference of confinement on the same droplet crossing the junction. Both conditions together create the necessary capillary pressure imbalance that eventually drives the lateral breakup. 

\subsection{Modelling the dynamics of the lateral breakup for lower Ca}

\begin{figure*}
\centering
  \includegraphics[width=0.9\linewidth]{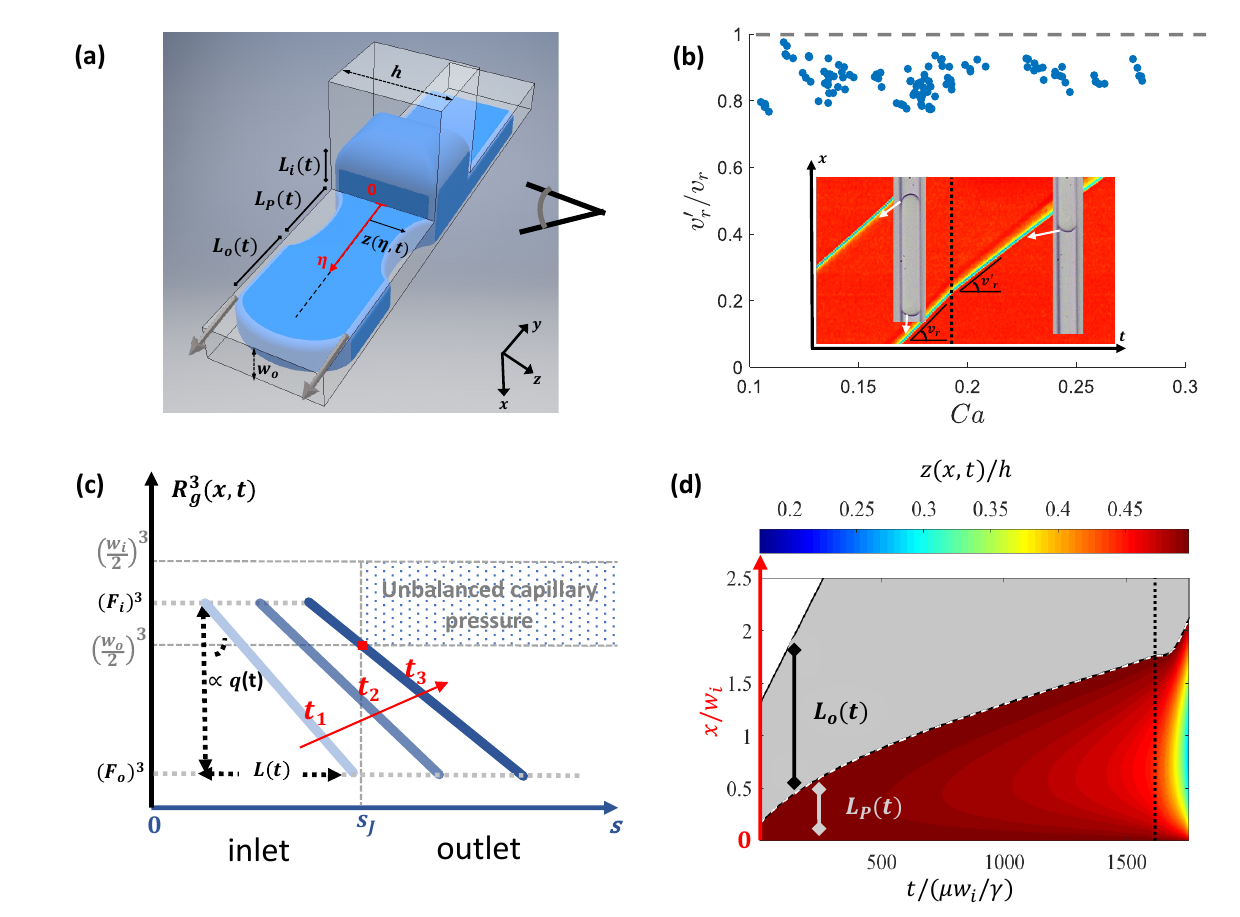}
\caption{\textbf{(a)} Modeling of the pocket development for one arm of the T junction, with the eye showing the observation perspective during experiment. During the inflation of the lateral pocket, the droplet can be divided into three regions with a length of $L_i(t)$,$L_p(t)$ and $L_o(t)$. The coordinate $\eta$ starts from the junction and is parallel to the outlet channel. In the parts of length $L_i(t)$ and $L_o(t)$, gutters are maintained, and the gutter flows are represented by grey arrows. \textbf{(b)} Relative rear cap velocity $v^{'}_{r} /v_{r}$ (see definition in the inset) once the droplet has entered the junction versus the droplet capillary number $Ca$ before having penetrated the junction. Inset shows the kymograph from which droplet cap trajectory is measured (yellow line), whose slope before and after the vertical dotted line gives $v_r$ and  $v^{'}_{r}$ respectively. \textbf{(c)} Result of step 1: schematic plot of equation (\ref{eq:gutter}) using junction coordinate S. Each blue curve represents the cubic power of the gutter radius along the droplet, from $F_i$ at the rear interface in the inlet to $F_o$ at the front interface in the outlet. Three such gutter radius profiles are shown corresponding to three time points, i.e., three droplet locations, as the droplet is advancing inside the channel. At t1: the droplet is about to enter the junction; t2: part of the droplet passes the junction, but the gutter radius inside the outlet channel is below $R_g^*=w_o/2$; t3: the droplet further advances, and $R_g=R_g^*$ is met at the red square; from this stage on, further advancing will cause $R_g>R_g^*$ inside the outlet channel, onset of necking, a pocket forms and inflates.\textbf{(d)} Result of step 2: temporal evolution of the pocket predicted by the model (for geometry $w_o/h=14/62$, $w_i/h=30/62$, with initial droplet length of $L/w_o = 8$ and capillary number of $Ca = 0.005$). The vertical axis corresponds to the $\eta$ coordinate (non-dimensionalized with $w_i$), and the length of $L_p(t)$ and $L_o(t)$ are drawn in color and grey respectively as a function of non-dimensional time (horizontal axis); The colormap represents the normalized local pocket depth $z(s,t)/h$ with the simulation stopped when $z(s,t)/h$ goes below $(w_o/h)/2 \approx 0.11$. 
\label{fgr:model}}

\end{figure*}
Next, we derive a theoretical model to describe the dynamics of the lateral breakup, occurring in two steps: 1. The droplet progresses through the channel until the necking criterion is met- this is the onset of the necking; and 2. The necking-induced pocket of continuous phase, fed via the gutters, inflates and thins down the droplet until final pinch-off. The first step is to find the condition where a minimal value of $1/R_g^*=2/w_o$ is met within the outlet channel. At an initial configuration, a droplet of speed Ca and an initial length $L(0)$ (not shown) passes the junction, creating gutters of length $L_i$ and $L_o$ in the inlet and outlet channel. We assume a homogeneous pressure $p_d$ inside the droplet, and obtain a hydraulic resistance per unit length:
\begin{equation}
r_h(s,t) = \frac{C \mu}{(1-\pi/4) R_g(s,t)^4},
\label{eq:gutres}
\end{equation}
where $\mu$ is the viscosity of the fluid, $C$ is a geometric constant with $C=93.93$ \cite{Ransohoff88}, and $R_g(s,t)$ is the radius of the gutter along the droplet internal coordinate $s$. Note that $s$ is oriented along $x$ in the inlet channel and along $y$ in the right outlet channel. 
At the rear and front caps we have $R_g(0,t)=F_i$ and $R_g(L(t),t) = F_o$, where $F_{i,o}=\kappa_{i,o}^{-1}$ is the inverse of the total curvatures of the caps defined in eq. (\ref{eq:kappa}). The difference between the two induces a flow rate $q(t)$ of the continuous phase, allocated in 4 gutters in the inlet channel and in $2 \times 2$ gutters in the outlet channels. Experimentally, we observed a reduction of droplet rear cap speed after the front cap enters the outlet channel (\textbf{Figure \ref{fgr:model}b}), which confirms the presence of this total bypass flux $q(t)$. 
Combining the Young-Laplace equation for the pressure balance at the interface $p(s,t) = p_d - \gamma/R_g(s,t)$ and the Poiseuille equation expressing the pressure gradient within the continuous phase $\partial p/\partial s = - r_h q /4$, one obtains an equation controlling the shape of the gutters:
\begin{equation}
\frac{\partial R_g}{\partial s}(s,t) = -\frac{q(t)}{4\gamma}\frac{C\mu}{(1-\pi/4)}\frac{1}{R_g(s,t)^2},
\label{eq:dRds}
\end{equation}
with a continuous change of gutter radius $R_g(s,t)$ along the droplet from $F_i$ to $F_o$. The quantity $L(t)= L_i(t)+L_o(t)$ designates the total length of the droplet (excluding caps) along $s$ the internal abscissa, and is determined by volume conservation from known initial droplet length $L(0)$ and $L_i(t)$ (see Appendix). Solving equation (\ref{eq:dRds}) provides both $q(t)$ -constant in space due to flow rate conservation- and the gutter radius $R_g(s,t)$ at any location ($s$) along the droplet:
\begin{equation}
q(t) =  \frac{\left ( F_i^3 - F_o^3 \right)}{A L(t)} ,\quad \text{and} 
\end{equation}

\begin{equation}
\quad \quad R_g^3(s,t) = F_i^3 - q(t)A s \quad \text{for}  \quad 0 \leq s \leq L(t),
\label{eq:gutter}
\end{equation}
where $A = 3C\mu/(4\gamma(1-\pi/4))$.  By progressively decreasing $L_i$ from $L(0)$ (the droplet turns the junction at t=0), we find the critical $L_{ic}$ such that $R_g = R_{g}^* = w_o/2$. Further advancing the droplet, $R_g > w_o/2$ cannot be met, and the necking has to start. The corresponding critical time $t_c$ can be obtained from $L_{ic} = L(0) - Ca\: t_c$. \textbf{Figure  \ref{fgr:model}c} plots  $R_g^3$, changing linearly from $F_i^{3}$ to $F_o^{3}$ along the droplet, for three time-stamps and corresponding droplet locations during the advancing of the droplet. It illustrates the following scenario for the onset of the necking: with the droplet advancing in the channel, the rear cap approaches the junction and increases more and more the gutter radius in the outlet channel, which eventually goes beyond the maximum possible value fixed by the outlet channel geometry, thus falling out of balance. Such process is strongly influenced by the flow rate and droplet size which change the slope and length of the $R_g^3$ curve. As the maximum gutter radius along the outlet channel is always attained at the junction location ($s_J$), it is always at the junction that the necking requirement is first met. Thus, the pocket of continuous phase is expected to start forming from the junction, which is confirmed by the experiment, as discussed below.



Now, the droplet enters a second phase consisting in the development of the pocket. To model the evolution of such process, we divide the droplet into three consecutive parts (\textbf{Figure \ref{fgr:model}a}): the part in the inlet channel of length $L_i(t)$ with gutters; The part on the spatial interval of $0\leq \eta \leq L_p$, where $L_p(t)$ is the length of the pocket, from the junction $(\eta=0)$ to where the droplet curvature in the direction of the flow vanishes $(\eta=L_p(t))$. The third part corresponds to the remaining of the droplet in the outlet channels where the gutter is resumed, of length $L_o(t)$. In the pocket region, we parametrize the droplet surface by its curvature in the flow direction (i.e, the curvature in the $y-z$ plane in \textbf{Figure \ref{fgr:model}a}), defined as 
\begin{equation}
k(\eta,t) = - \frac{\partial}{\partial \eta}\left[ \frac{\partial z(\eta,t)/\partial \eta}{\sqrt{1+\left(\partial z(\eta,t)/\partial \eta \right)^2}} \right].
\label{eq:curv}
\end{equation}
 Where $z\in[w_o/2;h/2]$ designates the interfacial position: $z$ equals to $h/2$ when no pocket is formed and gutters are maintained, and $z= w_o/2$ corresponds to a cylindrical droplet cross section that represents the end point of the pocket development. The curvature in the direction perpendicular to the flow, i.e, the curvature in the $x-z$ plane in \textbf{ Figure \ref{fgr:model}a}, is assumed constant and equals to $2/w_o$, such that the total curvature of the droplet in the pocket region writes $k(\eta,t)+2/w_o$. We then define the continuity equation for the continuous phase as $\partial S /\partial t = (\partial q/\partial x)/2$, where $S[z(\eta,t)]= (\pi-4)w_o^2/4 + 2 z(\eta,t) w_o $ is the cross-sectional area of the discrete phase in one outlet channel, which is fed by two quarter-sections at the top wall with a flow rate of $q(\eta,t)/4$ in each. We apply the derivative to the equation of the pressure balance at the interface to obtain the flow rate
\begin{equation}
q(\eta,t) = \frac{\gamma}{r_{h,S}[z(\eta,t)]}\frac{\partial k(\eta,t)}{ \partial \eta},
\label{eq:flowrate}
\end{equation}
where $r_{h,S}[z(\eta,t)]$ designates the hydrodynamic resistance per unit length of a quarter-section, which reduces to equation  (\ref{eq:gutres}) for $z=h/2$. 
Injecting both the expression for $S[z(\eta,t)]$ and equation (\ref{eq:flowrate}) in the continuity equation leads to 
\begin{equation}
\frac{\partial z(\eta,t)}{\partial t} = \frac{4 \gamma}{w_o} \frac{\partial }{\partial \eta}\left( \frac{1}{r_{h,S}[z(\eta,t)]}\frac{\partial k(\eta,t)}{\partial \eta}\right),
\label{eq:zk}
\end{equation}
such that equation (\ref{eq:curv}) and equation (\ref{eq:zk}) constitute a system of two coupled equations for the two unknowns $z(\eta,t)$ and $k(\eta,t)$. It is subject to two boundary conditions for $z$: $z(0,t) = z(L_p(t),t) = h/2$, as well as two for $k:$ $k(0,t) = \kappa_J - 2/w_o $, and $k(L_p(t),t) = 0$, where $\kappa_J$ is the curvature of the inlet gutter at the junction, found by matching with the inlet gutter regime (see Appendix). As mentioned, the necking condition has already been met, thus $k(0,t) \leq 0$, resulting in the opening of the pocket; however, $k(\eta,t)$ must increase with $\eta$ until recovering $k(L_p(t),t) = 0$, as the hydrodynamic resistance induces a pressure drop of the continuous phase (see equation (\ref{eq:flowrate})). We used a finite element method to calculate the dynamics of the necking process numerically , i.e. $z(\eta,t)$ and $k(\eta,t)$. The boundary conditions, numerical discretization and nondimensionalization used for this purpose are detailed in the appendix and numerical methods section.. The result of the simulation resolves the pocket evolution process, which can be represented by the change of pocket length $L_p(t)$ and depth $z(\eta,t)$ (\textbf{Figure \ref{fgr:model}d}). It shows that these two quantities increase over time, with a monotonic downstream extension of the pocket boundary. As $z(\eta,t) = w_o/2$, a locally cylindrical cross-section is reached during the pocket evolution, which triggers a fast breakup controlled by surface tension. 
\begin{figure*}[ht!]
\centering
  \includegraphics[scale=0.8]{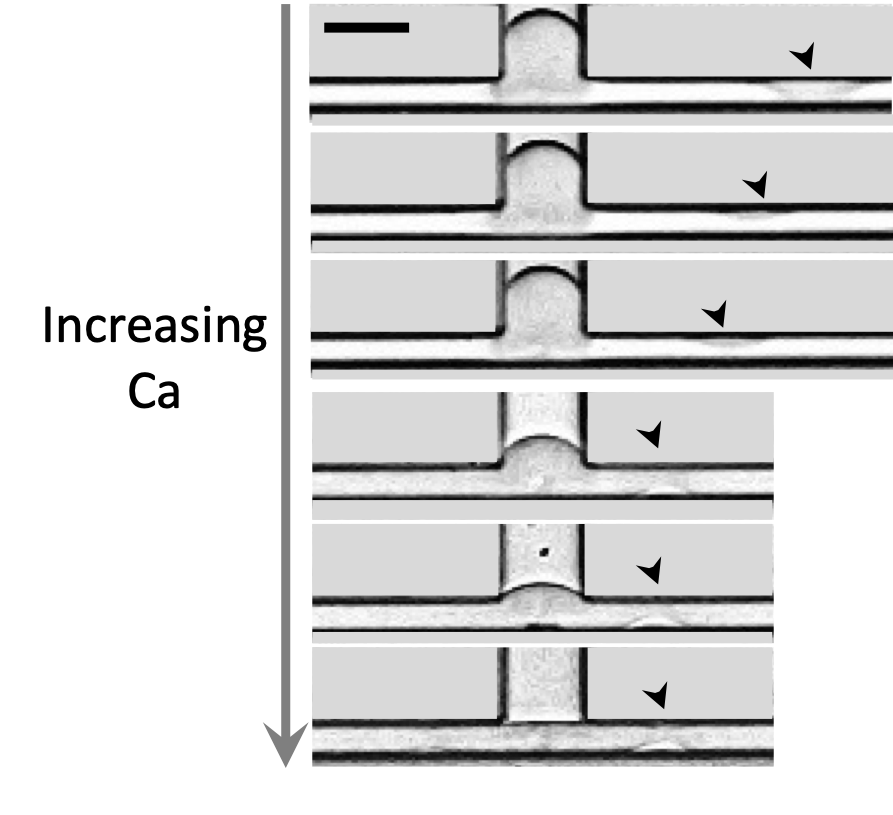}
\caption{Breakup moment of six breakup events with increasing $Ca$. The $Ca$ value from top to bottom is 0.016, 0.025, 0.056, 0.088, 0.129 and 0.177. The breakup location in the right outlet channel is shown with black arrowheads. The scale bar represents 30 $\mu$m.}
\label{fgr:obser}
\end{figure*}

\begin{figure*}[ht!]
\centering
  \includegraphics[scale=0.8]{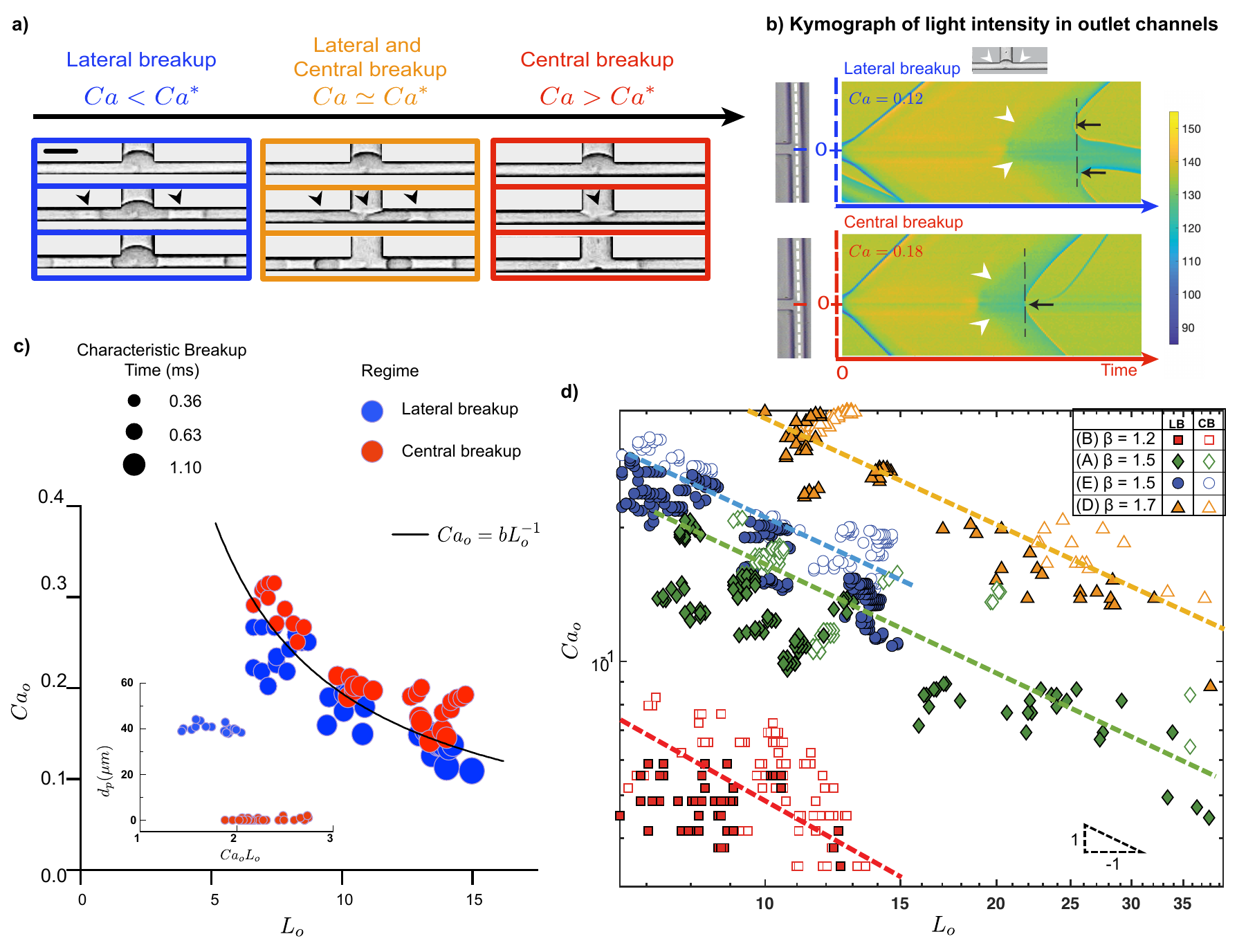}
\caption{\textbf{(a)} Time sequences of example breakup events for the same droplet length under three flow conditions. The breakup regime shifts from lateral to central breakup from low to high $Ca$. The scale bar represents 30 $\mu$m. \textbf{(b)} Kymograph of light intensity along the central part of the outlet channel (white dashed line), for two droplets of the same size but with different regimes; The grey value is turned into colors in the colormap. At time zero, both droplets enter the junction, the trajectory of the front interface forms a straight line. After the rear interface arrives at the junction (inset), the change of intensity due to the light scattering from the interface of the pocket is captured (white arrowheads), which always starts from the junction as predicted. At the end of each process, new interface(s) is formed at the locations indicated by the black arrows. \textbf{(c)} The breakup transition regime map of $Ca_o$ versus $\bar L_{o}$ for geometry A (\textbf{Figure \ref{fgr:static}c}), where $Ca_{o}=Ca(w_{i}/w_{o})/2$ and $\bar L_{o} = L(0)(w_{i}/w_{o})/w_{o}/2$. Blue and red circles represent lateral and central breakups respectively. The area of each circle is proportional to the characteristic breakup time, defined from when droplet rear interface reaches the junction to the final breakup moment. The black curve represents the function $Ca_{o} = b \bar L_{o}^{-1}$, where $b=1.9$ in this case (geometry A). Inset: $d_p$, the pinch off lateral distance from the junction center (in um) versus the product of $Ca_{o}\bar L_{o}$. \textbf{(d)} Logarithmic representation of the breakup transition regime map for $Ca_o$ vs. $\bar L_{o}$ for four geometries (A, B, D, E) with different $\beta$ values, represented by four colors. The filled and hollow markers represent lateral (LB) and central breakups (CB) respectively. Dashed lines of slope $-1$ are represented to guide the eyes.}
\label{fgr:tran}
\end{figure*}

This model is restricted to low $Ca$ for two main reasons. First, large values of $Ca$ may lead to a cross-section occupancy of both liquids which is not accounted for in our gutter model, as evidenced by Lozar \textit{et al.} \cite{Lozar07}. Second, the internal viscous dissipation has an increasing contribution at higher $Ca$, and was not included in the model.  Experimentally, we observed an evolution of lateral breakup behaviour from low to high $Ca$ condition. For lower $Ca$, a pocket forming process starts and stops before the rear interface reaches the junction. The corresponding breakup distance from the junction decreases with increasing $Ca$ (\textbf{Figure \ref{fgr:obser}}). For higher $Ca$, the breakup location is stabilized, and an decreasing rear cap curvature is observed as $Ca$ increases (\textbf{Figure \ref{fgr:obser}}). We next discuss the latter case concerning higher $Ca$.

\subsection{A central breakup recovered at higher $Ca$}

At higher $Ca$, when a critical $Ca^\ast$ is exceeded, the conventional central breakup regime can be recovered, even for geometries enabling lateral breakup. \textbf{Figure \ref{fgr:tran}a} shows three breakup events with the same droplet size. Central breakup is observed at higher values of $Ca$, while both lateral and central breakups can occur simultaneously near the critical value $Ca^\ast$. In \textbf{Figure \ref{fgr:tran}b} we show light intensity kymographs of two breakup events with the same droplet size but different breakup regimes. The existence of the lateral pocket cannot be imaged directly but it can be detected by a faint intensity change, caused by the  light scattering at the openings. First, it confirms that the necking starts from the junction as predicted by our model (\textbf{Figure \ref{fgr:model}c}). Second, it is remarkable to find out that the pocket formation occurs regardless of the final breakup outcome. This observation, together with the coexistence of lateral and central breakups near the critical $Ca^*$ indicates that the two processes are simultaneous. It gives a hint on the breakup transition mechanism, attributed to the faster completion of central breakup that aborts the lateral breakup process with the interfacial rearrangement.

We obtained the regime map near the transition zone for $Ca_o$ versus $\bar L_{o}$ in \textbf{Figure \ref{fgr:tran}c}. Here, $Ca_{o}=
Ca(w_{i}/w_{o})/2$ is the outlet channel capillary number, and $\bar L_{o} = L(0)(w_{i}/w_{o})/w_{o}/2$ is the initial droplet length translated into outlet channel (divided by 2 for only one branch) normalized by the outlet width. We compare the characteristic breakup time for both regimes, defined as the duration from the rear cap reaching the corner until the breakup and represented in Figure \ref{fgr:tran}c) as the surface area of the round markers. Interestingly, the characteristic time decreases approximately with increasing $Ca$ for both regimes. But at each transition point the central breakup always has a shorter characteristic time than the adjacent lateral breakup. From the regime map, a longer droplet needs a lower critical $Ca_o^*$ for the transition to central breakup. We found that the relation between $\bar L_{o}$ and its transition $Ca_o^*$ can be well described by the scaling law $Ca_o^{*}\sim b \bar L_{o}^{-1}$. The latter is also found to describe the non-breakup/central breakup transition on conventional T-junctions for long droplets \cite{Haringa2019}. This indicates that the lateral/central breakup transition is probably dictated by the enabling of the central breakup that is the faster process. This leads to a critical constant $(Ca_{o}\bar L_{o})^*$ that solely governs the lateral/central breakup transition. In the inset of \textbf{Figure \ref{fgr:tran}c}, the separation of the two regimes by $(Ca_{o}L_{o})^* \approx 1.9$ is shown.

  However, this experimentally determined prefactor is much higher than the theoretical value obtained for the same geometry (A) but assuming no lateral breakup following the scaling analysis of Haringa \textit{et al.} \cite{Haringa2019}(see Appendix). We note the difference of enabling central breakup on these new T-junctions. The continuous flow arriving at the junction can: (a) bypass the entire droplet through gutters, (b) flows into the lateral pocket and increase its volume, or (c) push the rear cap and contribute to central breakup. In conventional T-junctions, there is no pathway (b), and (a) is negligible at higher $Ca$\cite{Haringa2019}. In the lateral breakup-enabled T-junctions, the pathway (a) may be enhanced by an enlarged ``gutter'' area of the continuous phase when $Ca$ is high (prominent for high aspect ratio channel \cite{Lozar07}). Together with the uniquely presented pathway (b), it indicates that the lateral breakup-enabled T-junction uses a smaller portion of the incoming flow for central breakup compared to a conventional T-junction. This means for the same droplet length $\bar L_{o}$, a higher $Ca_o^*$ is required to enable the central breakup and the transition, reflected by the increased constant $(Ca_{o}\bar L_{o})^*$. In \textbf{Figure \ref{fgr:tran}c} we maps the breakup regimes near the transition zone for four geometries with the same outlet widths in a logarithmic scale (From geometries A,B,D and E described in the Table). The decay rate of $Ca_{o}^*$ with $\bar L_{o}$ is similar for all the tested geometries, as a power law with an exponent close to -1, similar to the above observation. The intersect of these -1 laws, proportional to the constant $(Ca_{o}L_{o})^*$, varies among the geometries. Geometries with more prominent lateral breakup (higher value of $\beta$) require a higher $Ca_{o}^*$ to recover central breakup. 
  
\begin{figure*}[ht!]
\centering
  \includegraphics[scale=0.75]{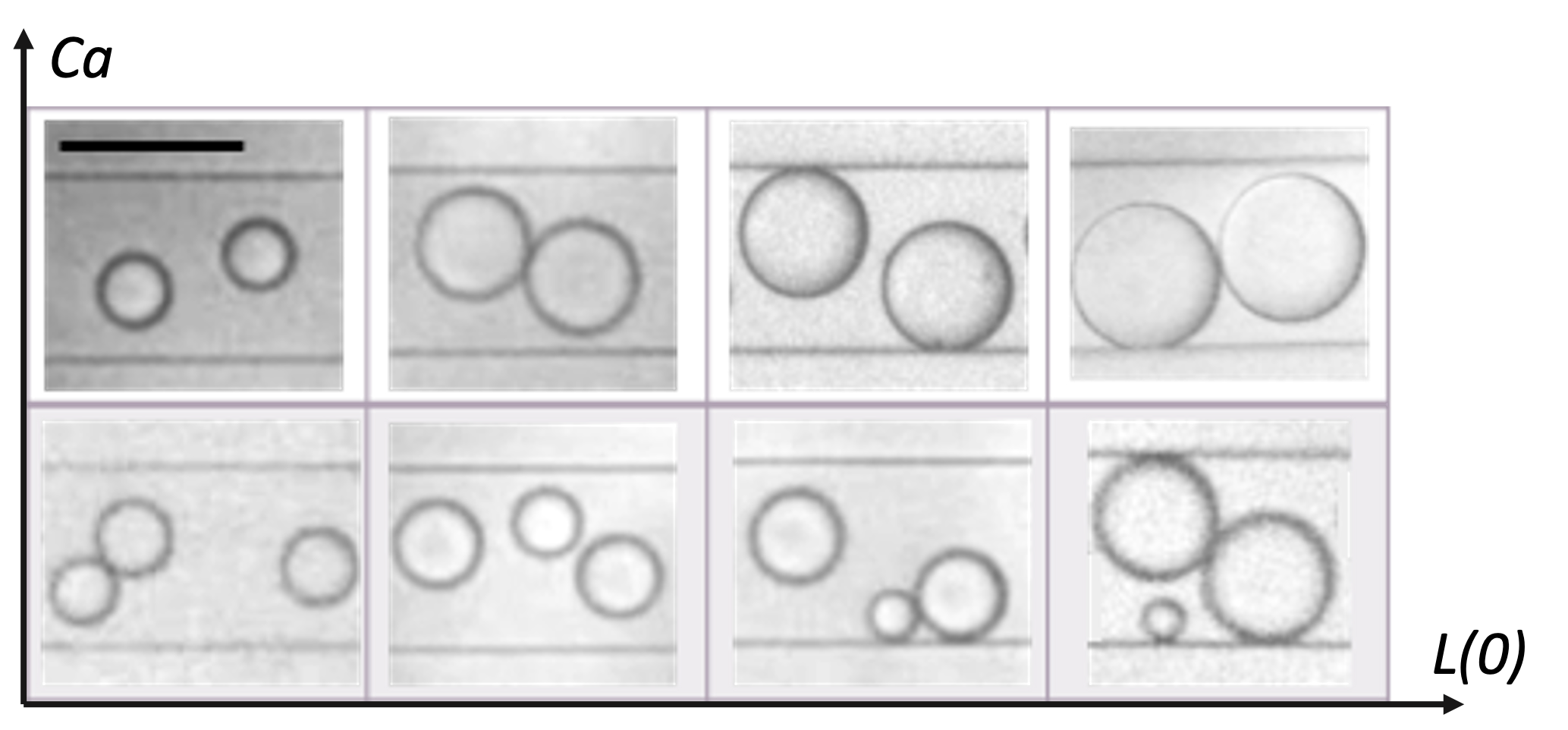}
\caption{Different daughter droplet composition from each droplet breakup event, with geometry B. From top to bottom: increasing the number of daughter droplets (by decreasing the capillary number $Ca$). From left to right: increasing the daughter droplet size and/or size ratio (by increasing the droplet length $L_o$). The scale bar represents 100 $\mu$m.}  
\label{fgr:bkr}
\end{figure*}

  We also mention that the geometry may alter the relative time scale of the two processes. For example, on geometries that favor less the lateral breakup (e.g. geometry B), there is no lateral/central simultaneous breakup near the transition region, as shown in \textbf{Figure \ref{fgr:tran}a} for geometry A, indicating a much slower lateral breakup at this moment than the central breakup. We thus warn that it is not impossible to have a lateral breakup process terminating before the central breakup process (of droplet flattening, concaving and finally pinching off), provided that a geometry promotes the former to have its process finished earlier. In a nutshell, the droplet breakup fate in these novel T-junctions should be determined by the temporal dynamics of both breakup processes and is dominated by the faster one. This means the two most common droplet formation mechanisms in microfluidics- driven by hydrodynamic force and driven by surface tension, simultaneously compete in the same geometry. By merely shifting flow condition, the droplet size and/or composition can be changed on-fly (\textbf{Figure \ref{fgr:bkr}}). 

\section{Conclusion and outlook}

In summary, we reported on a novel lateral droplet breakup occurring in microfluidic T-junctions which leads to the formation of three daughter droplets. We experimentally evidenced that this new regime arises from an unbalanced capillary pressure at the drop interface induced by the strong gradient of confinement across the junction (provided that $h>w_i>w_o$). A geometrical design rule was proposed accordingly to enable the lateral breakup regime. We also developed a model depicting the development of the lateral pockets responsible for the ultimate lateral breakup, for low capillary number $Ca$. Furthermore, we showed that a unique central breakup is recovered at higher $Ca$, a mechanism observed in conventional T-junction. We showed that the critical capillary number $Ca^\ast$ marking this transition from lateral to central breakup is compatible with an inverse dependency on the droplet length. The presence of the lateral pockets and their inflation explains that the values of $Ca^\ast$ are orders of magnitude higher than predicted by a scaling analysis in the spirit of Haringa \textit{et al\cite{Haringa2019}}. Accounting for a thickening of the gutter at higher $Ca$, as observed by Lozar \textit{et al.} \cite{Lozar07}, was not sufficient to explain high values of $Ca^\ast$ reported. A thorough theoretical determination of $Ca^\ast$ remains to be achieved and is a challenge for future studies.

In the end, both hydrodynamic-force-driven and surface-tension-driven mechanisms are enabled on one geometry which allows new microfluidics functionalities. On one hand, active control over the flow condition can change droplet composition and sizes without changing geometries; On the other hand, without active control, the mere change of the content or property of the droplet could possibly alter the temporal competition of the two breakup regimes thus shifting the breakup results for passive applications. Altogether, we expect that these new breakup phenomena will provide versatile tools to the community to manipulate and control the volume of droplets.

\section*{APPENDIX A - Mathematical modelling of the inflation of the lateral pockets at low $Ca$}
\label{SI:math}

Hereafter, we fully characterize the inlet and outlet gutters regime of the droplet (shown in \textbf{Figure \ref{fgr:model}a}), as they couple with equation (\ref{eq:curv}) and (\ref{eq:zk}) for the pocket dynamic trough the boundary conditions and the conservation of the droplet volume. We then show that they constitute all together a closed system that is made non-dimensional, and after some mathematical rearrangements, can be directly integrated in time.

Let $R_g^{(i)}(s,t)$ designate the radius of the gutter in the inlet channel (see \textbf{Figure \ref{fgr:model}a}) whose internal coordinate is $s \in [0;L_i(t)]$, purely along the $x$ direction. Solving eq.(\ref{eq:dRds}) subject to $R_g^{(i)}(0,t)=F_i$ leads to 

\begin{equation}
R_g^{(i)}(s,t)^3 = F_i^3-q_i(t)As
\end{equation}

where $q_i(t)$ is the inlet gutter flow rate. By continuity, we can express it in terms of $k(\eta,t)$ using eq.(\ref{eq:flowrate}) as

\begin{equation}
q_i(t) = \frac{4 \gamma}{r_{h,S}[z(0,t)=h/2]}\frac{\partial k(0,t)}{\partial \eta},  
\end{equation}

\noindent (we recall that $\eta=0$ designates the opening of the pocket at the junction and $\eta=L_p(t)$ its closure (see \textbf{Figure \ref{fgr:model}a} ). The four boundary conditions for $z(\eta,t)$ and $k(\eta,t)$ write $z(0,t) = z(L_p(t),t) = h/2$, $k(0,t) = 1/R_g^{(i)}(L_i(t),t)-2/w_o$, and $k(L_p(t),t) = 0$. The outlet gutter, of length $L_o(t)$ (see \textbf{Figure \ref{fgr:model}a} ), has a radius $R_g^{(o)}(\cpo,t)$ whose internal coordinate is $\cpo \in[0;L_o(t)]$, purely along the $y$ direction. The radius $R_g^{(o)}(\cpo,t)$ is characterized by solving eq.(\ref{eq:dRds}) subject to $R_g^{(o)}(0,t)=w_o/2$ (since $k(L_p(t),0)=0$) : 
\begin{equation}
R_g^{(o)}(\cpo,t)^3 = \left(\frac{w_o}{2}\right)^3 -q_o(t) A \cpo, \quad \text{where} \quad q_o(t) = \frac{4 \gamma}{r_{h,S}[z(L_p(t),t)=h/2]}\frac{\partial k(L_p(t),t)}{\partial \eta} 
\end{equation}
\noindent Its length $L_o(t)$ is easily determined by imposing $R_g^{(o)}(L_o(t),t) = F_o$. As will become clear in a moment, the problem is closed by imposing of conservation of the total volume of the droplet. From now on quantities are made non-dimensional, by $h$ in space and $\mu h/\gamma$ in time
\begin{align*}
&h(\tilde{s},\tilde{\eta},\tilde{\cpo}) = (s,\eta,\cpo), \quad \frac{\mu h}{\gamma} \tilde{t} = t, \quad h^2 \tilde{S} = S, \quad \frac{h^2 \gamma}{\mu} \tilde{q} = q, \quad \frac{1}{h}\tilde{k} = k, \quad h \tilde{z} = z, \\ &\frac{\mu}{h^4} \tilde{r}_{h,S}  = r_{h,S}, \quad \text{and} \quad \frac{\gamma}{\mu} Ca = U.
\end{align*}
Equations (\ref{eq:curv}) and (\ref{eq:zk}) thus non-dimensionalized, they are then put in the form of a classical conservation law for $\tilde{z}(\tilde{\eta},\tilde{t})$ and $\tilde{k}(\tilde{\eta},\tilde{t})$:
\begin{align}
\begin{bmatrix}
1 & 0\\ 
0 & 0
\end{bmatrix} \frac{\partial }{\partial t }\begin{bmatrix}
z \\ 
k
\end{bmatrix} + \frac{\partial }{\partial \eta }\begin{bmatrix}
 -  B \frac{1}{r_{h,S}[z]} \frac{\partial k}{\partial \eta} \\ 
 - \frac{\partial z}{\partial \eta }\left[1+\left(\frac{\partial z}{\partial \eta}\right)^2 \right]^{-1/2} 
\end{bmatrix} = \begin{bmatrix}
0 \\ 
k
\end{bmatrix}, \quad \text{for}\quad 0\leq \eta \leq L_p(t)
\label{eq:sys}
\end{align}

\noindent where all the tildes have been dropped and $B = 1/w_o$ is a constant (it is understood that $L_p$ and $w_o$ have been made non-dimensional by $h$, etc...). System (\ref{eq:sys}) is re-written under the change of variable $\xi = \eta/L_p(t) $ in order to be solved over the time-independent domain $0 \leq \xi \leq 1$, which is significantly more convenient. The partial derivatives are transformed as $\partial_t \rightarrow \partial_{t} - (\mathrm{d}_t L_p)L_p^{-1}  \xi \partial_{\xi}$ and $\partial_{\eta} \rightarrow L_p^{-1}\partial_{\xi}$:

\begin{align}
\begin{bmatrix}
1 & 0\\ 
0 & 0
\end{bmatrix} \frac{\partial }{\partial t}\begin{bmatrix}
\bar{z} \\ 
\bar{k}
\end{bmatrix} + \frac{\partial }{\partial \xi }\begin{bmatrix}
 - \frac{\mathrm{d} L_p}{\mathrm{d} t} \frac{1}{L_p} \xi \bar{z} -  \frac{B}{L_p^2}\frac{1}{r_{h,S}[\bar{z}]} \frac{\partial \bar{k}}{\partial \xi} \\ 
 - \frac{1}{L_p^2}\frac{\partial \bar{z} }{\partial \xi} \left[1+\frac{1}{L_p^2}\left( \frac{\partial \bar{z} }{\partial \xi}  \right)^2 \right]^{-1/2}   
\end{bmatrix} = \begin{bmatrix}
- \frac{\mathrm{d} L_p}{\mathrm{d} t} \frac{1}{L_p} \bar{z}  \\ 
\bar{k}
\end{bmatrix}
\label{eq:sysnv}
\end{align}
where $z(\eta,t) = \bar{z}(\xi,t)$, then $k(\eta,t) = \bar{k}(\xi,t)$. Under this change of variables the flow rate $q(\eta,t) = \bar{q}(\xi,t)$ occurring inside the pocket, and the flow rates $q_i(t)$ and $q_o(t)$ occurring inside the inlet and outlet gutters, respectively, express :
 \begin{align*}
\bar{q}(\xi,t)  &= \frac{4 }{r_{h,S}[\bar{z}(\xi,t)] L_p(t) }\frac{\partial \bar{k}}{\partial  \xi}(\xi,t),
\quad
q_i(t) = \bar{q}(0,t)\quad, 
\text{and}  
\quad
q_o(t) = \bar{q}(1,t).
\end{align*}
The boundary conditions of system (\ref{eq:sysnv}) are re-written : 
\begin{equation}
\begin{split}
&\bar{z}(0,t) = 1/2, \quad \bar{z}(1,t)  = 1/2, \\ 
&\bar{k}(0,t) = \left[\left( F_i^3 - q_i(t) A L_i(t) \right)^{-1/3} -   \frac{2}{w_o}  \right], \quad \text{and} \quad \bar{k}(1,t) = 0.
\end{split}
\label{eq:bcc}
\end{equation}

\noindent(with the non-dimensional $A=3C/(4-\pi)$). A third equation is necessary for the third unknown $L_p(t)$, and the problem is closed by imposing the volume conservation $V(t) = V(0)$. Let $V_g^{(i)}(t)$ designate the volume of the part of the droplet contained in the inlet channel and where a gutter is present (i.e. for $0\leq s \leq L_i(t)$) ; the volume of the rear cap is in addition $V_{cap}^{(i)}(t)$. Accordingly, let $V_g^{(o)}(t)$ be the volume of the part of the droplet contained in one of the two outlet channels and where a gutter is present (i.e. for $0\leq \cpo \leq L_o(t)$) ; the volume of one front cap is in addition $V_{cap}^{(o)}$. We recall that the expression for the cross sectional area of the discrete phase in the pocket region is $S[z(\eta,t)] = (\pi-4)w_o^2/4 + 2 w_o z(\eta,t)$, such that it is associated to a volume contribution in a outlet channel of $\int_{0}^{\eta=L_p(t)}S[z(\eta,t)]d\eta$. Eventually : 
\begin{multline}
V(t) =  2V_{cap}^{(o)} + V_{cap}^{(i)}(t) + V_g^{(i)}(t) + 2V_g^{(o)}(t) + 2\int_{0}^{\eta=L_p(t)}S[z(\eta,t)]d\eta \\ 
= 2V_{cap}^{(o)} + V_{cap}^{(i)}(t) + V_g^{(i)}(t) + 2V_g^{(o)}(t) + 2 L_p(t) \left[ \frac{(\pi-4) w_o^2}{4} + 2 w_o \int_{0}^{\xi = 1} \bar{z}(\xi,t) d\xi \right] = V(0)  \label{eq:forLp}
\end{multline}
In addition, the cross sectional area the drop in a region where a gutter (of radius $R_g$) is present is $S_g^{({i,o})} = h w_{i,o} - \left( R_g^{({i,o})} \right)^2(4-\pi)$, such that

\begin{align*}
 V_g^{(o)}(t) &= \int_{\cpo=0}^{\cpo=L_o(t)} S_g^{(o)}(\cpo,t) d\cpo = \frac{1}{A q_o(t)} \left[  h w_o \left(\left(\frac{w_o}{2}\right)^3-F_o^3 \right)- \frac{3(4-\pi)}{5} \left(\left(\frac{w_o}{2}\right)^5 -F_o^5 \right)\right], 
\end{align*}
and we compute similarly : 
 \begin{align*}
V^{(i)}_{g}(t) = h w_i L_i(t)- \frac{3(4-\pi)}{5 A q_i(t)} \left[ F_i^5 - ( F_i^3 - L_i(t) A q_i(t))^{5/3} \right]
\end{align*}

In order to mimic a constant-velocity progression of the droplet in the inlet channel before the rear droplet interface reaches the junction, as observed experimentally, the length $L_i(t)$ is chosen as a ramp in time
\begin{equation}
L_i(t) = \begin{cases}
L_i(0)-0.85 \cdot t \cdot Ca & \text{ if } 0 \leq t \leq t_J  \\ 
0 & \text{ if } t \geq t_J 
\end{cases}
\label{eq:ramp}
\end{equation}
which acts as a source of excitation for the system (\ref{eq:sysnv}) trough the boundary condition for $\bar{k}(0,t)$ in (\ref{eq:bcc}). At the time $t_J = L_i(0)/(0.85 Ca)$ the rear droplet interface reaches the junction, and $L_i(0)$ is such that the necking condition as shown in \textbf{Figure \ref{fgr:model}c} is met. The coefficient $0.85$ multiplying $Ca$ arises from the experimental data (shown in \textbf{Figure \ref{fgr:model}b}). The rear cap volume and curvature need also to be treated differently depending on whether $t$ is smaller or larger than $t_J$; let $V_{cap}[h,w_i]$ designate the equilibrium volume of the rear cap, whose value is taken from Musterd \textit{et al.}\cite{Musterd15} for $H/W \leq 1$
\begin{align*}
A_{bd}[H,W] &= HW - 4\cdot F[W,H]^2(1-\pi/4)  = HW +  F[W,H]^2(\pi-4) \\
L_{cap}[H,W]&=   \frac{W}{2} \\
V_{cap}[H,W]&=  \int_{0}^{L_{cap}} A_{bd}[H,W] \left( 1 -\frac{y^2}{L_{cap}[H,W]^2} \right ) dy = \frac{2}{3} L_{cap}[H,W]A_{bd}[H,W]  \\
\end{align*}
Thereby the rear cap volume $\Vic(t)$ is implemented as
\begin{equation}
\Vic(t) = \begin{cases}
V_{cap}[h,w_i]=cst & \text{ if } 0 \leq t \leq t_J  \\ 
V_{cap}[h,w_i]\frac{\mathcal{C}i(t)}{\kappa_i} & \text{ if } t \geq t_J 
\end{cases}
\label{eq:Vict}
\end{equation},
where $\mathcal{C}i(t)$ is the rear cap curvature with $\mathcal{C}i(t\leq t_J) = \kappa_i$. Above $t=t_J$, the rear droplet interface must undergo the incoming flow rate $h w_i Ca$ and the following equation for the cap volume is activated
\begin{equation}
\frac{\mathrm{d} \Vic}{\mathrm{d} t} = q_i(t) - Ca \:h w_i
\label{eq:Vci}
\end{equation}
Injecting (\ref{eq:Vict}) in (\ref{eq:Vci}) leads to an evolution equation for $\mathcal{C}_i(t)$ for $t \geq t_J$, and the boundary condition for the curvature at the opening of the pocket is accordingly replaced by $k(0,t) = \mathcal{C}_i(t)-2/w_o$ for these times.

\section*{APPENDIX B - A scaling argument for onset of central breakup}

In the spirit of reference\cite{Haringa2019}, we try to obtain a scaling of the critical capillary number for central breakup to happen. Neglecting the presence of a lateral inflating pocket, the inversely proportional relationship between $C_a^*$ and $L_o/w_o$ might be understood as follows: after the body of the droplet has fully entered the outlet channels, we consider a threshold situation where the rear cap is flat for the observer, such that the rear cap curvature is assumed $1/F_i = 2/h$, although the precise value of the latter has little influence on the scaling law derived hereafter. The threshold capillary number above which a central breakup (CB) is expected is obtained by balancing the continuous flow rate arriving in the junction ($C_a w_i h$), with the capillary flow rate in the four gutters located in the outlet channel ($4(1/F_o-2/h)/(r_o L_o)$), the quantity $r_o$ being the hydraulic resistance per unit length of a gutter in the outlet channel. This leads to : 

\begin{equation}
\frac{w_o}{w_i}C_a^* \sim \frac{4 w_o }{ w_i^2 h r_o L_o}\left( \frac{1}{F_o}-\frac{2}{h}\right)  \propto \left(\frac{L_o}{w_o}\right)^{-1} 
\label{eq:Cat}
\end{equation}

For, $w_o=14/62$, $w_i=30/62$, $h=1$ and using eq.(\ref{eq:kappa}) for $F_o$ and eq.(\ref{eq:gutres}) with $R_g = w_o/2$ for $r_o$, we obtain $4\left( \frac{1}{F_o}-\frac{2}{h}\right)/(w_i^2 h r_o) \approx 5\cdot 10^{-5}$. However, this corresponds to a $w_o/w_i C_a^*$ approximately four orders of magnitude smaller than experimental values.

\bibliography{apssamp}

\end{document}